\documentstyle[12pt]{article}
\date{\today}

\def\be{\begin{equation}}
\def\ee{\end{equation}}
\def\bear{\begin{eqnarray}}
\def\eear{\end{eqnarray}}
\def\nn{\nonumber}

\def\half{{{1\over 2}}}

\newcommand\px[1]{{\partial_{#1}}}

\newcommand\rep[1]{{\underline{\bf {#1}}}}      % representation
\def\BC{{{\bf C}}}

\newcommand\MS[1]{{{\bf S}^{#1}}}               % Circle, sphere,...
\newcommand\MT[1]{{{\bf T}^{#1}}}               % Torus
\newcommand\CP[1]{{{\bf CP}^{#1}}}              % CP
\newcommand\SUSY[1]{{{\cal N}= {#1}}}           % N=? SUSY
\def\LeftPar{{\raisebox{-3pt}{\Huge (}}}        % Big left parenthesis
\def\RightPar{{\raisebox{-3pt}{\Huge )}}}       % Big right parenthesis

\def\a{{\alpha}}
\def\b{{\beta}}

\def\u{{\mu}}

\begin{document}
\begin{titlepage}
\titlepage
\rightline{PUPT-1896, NSF-ITP-99-147}
\rightline{hep-th/0002097}
\rightline{Feb 10, 2000}
\vskip 1cm
\centerline {{\Large \bf Chiral Compactifications
             of 6D Conformal Theories}}
\vskip 1cm

\centerline {Chang S. Chan\footnote{cschan@princeton.edu},
    Ori J. Ganor\footnote{origa@viper.princeton.edu}}

\vskip .5cm

\begin{center}
\em  Department of Physics \\
Jadwin Hall \\
Princeton University\\
NJ 08544
\end{center}
\vskip 1cm

\centerline{and}
\centerline{Morten Krogh\footnote{krogh@itp.ucsb.edu}}
\vskip .5cm
\begin{center}
\em  Institute for Theoretical Physics\\
University of California, Santa Barbara, CA 93106 \\
\end{center}

\vskip 1cm

\abstract{
We construct chiral $N=1$ gauge theories in 4D by compactifying
the 6D Blum-Intriligator $(1,0)$ theories of 5-branes at $A_k$
singularities on $T^2$ with a nontrivial bundle of the global $U(1)$
symmetry of these theories.
}
\end{titlepage}
\tableofcontents
%\vskip 0.5cm
%{\bf this is file ``intro.tex''}
%\vskip 0.5cm

\section{Introduction}
% =======================================================================
In recent years it has been realized that many 3+1D gauge theories
can be obtained as special low-energy limits of compactified 5+1D
superconformal theories. Some of the known 5+1D theories are the
$\SUSY{(2,0)}$ theory \cite{WitCOM}, the $E_8$ $\SUSY{(1,0)}$
theory~\cite{GanHan} and the Blum-Intriligator~(BI) \cite{BluInt}
theories of $N$ M5-branes at an $A_{k-1}$ singularity.

Indeed,
part of the appeal of these theories is that by compactification
on $\MT{2}$ we can get various gauge theories in 3+1D at low-energy.
Thus, $\SUSY{4}$ SYM is obtained from the (2,0)-theory~\cite{WitCOM}
and $\SUSY{2}$ SYM with various matter content is obtained
from the $E_8$ $\SUSY{(1,0)}$ theory~\cite{GanCOM,GMS}.

Starting with the 5+1D BI theories we can compactify on $\MS{1}$
to get, at low-energies, the $\SUSY{2}$ quiver gauge theories
with gauge group $SU(N)^k$ and bi-fundamental matter
hypermultiplets \cite{DouMoo}. One can also realize a mass to the
hypermultiplets by using the global $U(1)$ symmetry of the BI
theories. Turning on a small background Wilson line for that
$U(1)$ corresponds at low energy to turning on the mass \cite{CGKM}.

In this paper we will construct chiral 3+1D theories from the BI
theories.  As an intermediate step, we start with a 4+1D hypermultiplet.
Given a hypermultiplet in 4+1D we can construct a low-energy chiral
multiplet as follows. Let us take an infinite $5^{th}$ direction
and let us give the fermions of the hypermultiplet a mass $m(x_5)$
that varies along the $5^{th}$ direction from $m=-\infty$ at
$x_5=-\infty$ to $m=\infty$ at $x_5=\infty$ (see \cite{ADD}). As
we shall review below, if we also let the scalar fields have
masses $\sqrt{m^2 \pm {{dm}\over {dx_5}}}$ then in the remaining 4
dimensions $\SUSY{1}$ supersymmetry is preserved and at low
energies we get a chiral multiplet localized near the point where
$m(x_5)=0$.
Thus, by varying the mass of the hypermultiplets in a 5D
gauge theory along the $5^{th}$ direction, we can obtain, at low-energies,
a chiral gauge theory in 4D.

A 5D gauge theory is only defined as a low-energy effective
action. However, we can realize it as a 6D theory compactified
on a circle. We would like to elevate the construction
of chiral gauge theories to 6D.
One motivation for that is that a 6D realization often provides
insight into the strong coupling behavior of the theory. 
The 6D theories that we will use are the BI theories and the
construction of chiral gauge theories from their compactifications
is the purpose of this paper.

The paper is organized as follows. In section (2) we review the
example of a 4+1D hypermultiplet. In section (3) we study the
compactification of a general 5+1D theory.  In section (4) we
discuss the BI theories and their compactification.

\section{A free hypermultiplet}
% =======================================================================
In this section we will study a free hypermultiplet in 5+1D and
4+1D. The reason for studying this simple system is that it gives
us an explicit realization of the mechanism which produces chiral
matter in 3+1D. We will later apply the same type of
compactification to obtain chiral matter in 3+1D starting from
5+1D theories.

We will show that a 4+1D hypermultiplet with a mass that varies
along the 5th direction preserves $\SUSY{1}$ SUSY in 3+1D and gives rise
to chiral multiplets. The
 4+1D hypermultiplet with a
 varying mass can be obtained from
a 5+1D hypermultiplet compactified on a circle
 and coupled to a background field.

\subsection{A 5+1D chiral hypermultiplet}
% -----------------------------------------------------------------------
A convenient way of getting the quantum numbers of a 5+1D hypermultiplet
is to start from 9+1D super Yang-Mills reduced to 5+1D.
This theory comprises of a
single multiplet under the $\SUSY{(1,1)}$ SUSY.
However, under an $\SUSY{(1,0)}$ subgroup of the supersymmetry algebra
it decomposes into a vector-multiplet and a hypermultiplet.
The statements below follow easily by thinking about the system
in this way.

A hypermultiplet in 5+1D (with $\SUSY{(1,0)}$ supersymmetry)
contains 4 real scalars and one chiral fermion.
It is convenient to decompose the components under the Lorentz group
$SO(5,1)$, the R-symmetry group $SU(2)_R$ and the global flavor
symmetry $SU(2)_F$.
Under $SO(5,1)\times SU(2)_R\times SU(2)_F$ the SUSY generators $Q_\a^i$
transform
as $(\rep{4},\rep{2},\rep{1})$. Note that both $\rep{4}$ and $\rep{2}$
are pseudo-real  representations so one can add a reality condition
to have 8 real SUSY generators.
Here $i=1,2$ is an index of the $\rep{2}$ of
$SU(2)_R$ and $\a=1\dots 4$ is an index of the $\rep{4}$ of $SO(5,1)$.
We will assume that the hypermultiplet is charged under $SU(2)_F$.
The fermions of the hypermultiplet transform as
$(\rep{4},\rep{1},\rep{2})$ with an added reality condition.
We will denote them by $\psi_\a^a$ with $a=1,2$ an index of $SU(2)_F$.
The bosons transform as $(\rep{1},\rep{2},\rep{2})$ and will be denoted
by $\phi^{i a}$.
 Recall that the Dirac matrices, $\Gamma^\u_{\a\b}$
($\u=0\dots 5$), of $SO(5,1)$ can be chosen to be anti-symmetric.
In the rest of the paper they will be anti-symmetric.
We will also use the anti-symmetric $\epsilon_{ij}$ of the $\rep{2}$
of $SU(2)_R$ to lower and raise the indices $i,j=1,2$.

The reality conditions are,
\bear
(\phi^{i a})^\dagger =
{C_b}^a {C_j}^i \phi^{j b},\qquad
(\psi_\b^b)^\dagger = {C_a}^b {C_\b}^\a\psi_\a^a, \label{reality}
\eear
where ${C_b}^a$, ${C_j}^i$
 and ${C_\b}^\a$ are the charge conjugation matrices
of (respectively) $\rep{2}$ of $SU(2)_F$, $\rep{2}$ of $SU(2)_R$
and $\rep{4}$ of $SO(5,1)$.

The action is
$$
S = \int d^6 x \left( - {1\over 4} \epsilon_{ij}\epsilon_{ab}
\partial_{\mu}\phi^{ia} \partial^{\mu}\phi^{jb}
+ \half \epsilon_{ab} \psi_\a^a \Gamma^{\mu \a \b}\partial_\u
\psi^b_\b \right).
$$
Our sign conventions are $\epsilon_{12}=\epsilon^{12}=1$.
The equations of motion derived from this
action are
$$
\Box\phi^{i a} = 0,\qquad \Gamma^{\u,{\a\b}}\px{\u}\psi_\a^a = 0.
$$
The supersymmetry transformations are:
$$
\delta\phi^{i a} =2 \eta^{\a i}\psi_\a^a,\qquad
\delta\psi_\a^a = \epsilon_{ij}\eta^{\b i}\Gamma^\u_{\a\b}\px{\u}\phi^{j a}.
$$

\subsection{A 4+1D massive hypermultiplet}\label{masshyp}
% -----------------------------------------------------------------------
Now we will consider a massive hypermultiplet in 4+1D.
The quantum numbers, action and supersymmetry
transformations of this can easily be obtained from the 5+1D case.
We consider a 5+1D hypermultiplet with a specific $x^5$ dependence.
\bear
\phi^a(x,x^5) &=& \phi^b(x){(e^{imx^5\tau^3})^a}_b \nn \\
\psi^a(x,x^5) &=& \psi^b(x){(e^{imx^5\tau^3})^a}_b \label{reduction}
\eear
Here $x$ stands for $x^0,x^1,x^2,x^3,x^4$ and
$$
\tau^3 = \left(\matrix{ 1 & 0 \cr 0 & -1 \cr} \right)
$$
is inserted to give the right sign in the exponential.
$\phi^1$ and $\phi^2$ must have different signs because of
the reality condition (\ref{reality}).
The quantum numbers are the same as in 5+1D.
A 4+1D massive hypermultiplet contains 4 bosons $\phi^{i a}$
in the $(\rep{1},\rep{2},\rep{2})$ of 
$SO(4,1)\times SU(2)_R\times SU(2)_F$,
where $SO(4,1)$ is the Lorentz group and $SU(2)_R$ and $SU(2)_F$
are the R-symmetry and flavor symmetry, respectively.
It also has fermions $\psi^{\a a}$ in the $(\rep{4},\rep{1},\rep{2})$.
Recall that the representation $\rep{4}$ of $SO(4,1)$ has an invariant
anti-symmetric form $\epsilon_{\a\b}$ which we will use to lower and raise
indices. From the 5+1D point of view that is just $\Gamma^5$ which
commutes with $SO(4,1)$ transformations.
The action in 4+1D is obtained simply by plugging the fields
in (\ref{reduction}) into the 5+1D
action.
$$
S = \int d^5 x \left( - {1\over 4} \epsilon_{ij}\epsilon_{ab}
(\partial_{\mu}\phi^{ia} \partial^{\mu}\phi^{jb}
+ m^2 \phi^{ia} \phi^{jb})
+ \half \epsilon_{ab} \psi_\a^a \Gamma^{\mu \a \b}\partial_\u
\psi^b_\b
+ \half i m \epsilon_{ab}{(\tau^3)_c}^b \psi_\a^a
 \Gamma^{5 \a \b}\partial_5 \psi^c_\b \right)
$$
The equations of motion follow:
$$
(\Box + m^2)\phi^{i a} = 0,\qquad
\Gamma^{\u,{\a\b}}\px{\u}\psi_\a^a +
 i m  {(\tau^3)_b}^a\Gamma^{5 \a\b}\psi_\a^b =0.
$$
The reality conditions on the fields are the same as in 5+1D,
as is obvious from the way we obtained them.
The SUSY transformations are obtained from the 5+1D
transformations:
\bear
\delta\phi^{i a} &=&2 \eta^{\a i}\psi_\a^a,\nn\\
\delta\psi_\a^a &=&
  \epsilon_{ij}\eta^{\b i} \Gamma^\u_{\a\b}\px{\u}\phi^{j a} 
 +im \epsilon_{ij}\eta^{\b i}\Gamma^5_{\a\b}(\tau^3)^a_b \phi^{j b}.
\label{transf}
\eear

\subsection{Variable mass}\label{varmass}
% -----------------------------------------------------------------------
We will now discuss a reduction of the 4+1D massive hypermultiplet
to 3+1D in a way that preserves half the supersymmetry (i.e.
$\SUSY{1}$ in 3+1D) and can produce chiral multiplets. This
reduction was also discussed in \cite{ADD}. We pick a spatial direction
$x^4$ and let the mass vary as a function of $x^4$ only. Let this
function be $m(x^4)$. In the previous subsection we wrote down the
action and supersymmetry transformations for a massive
hypermultiplet. The mass, $m$, was constant. The question is what
action should we use when $m$ is not constant. The
only condition the new action must fulfill is that it reduces to
the usual one when $m$ is constant. However that only determines
the action up to terms involving derivatives of $m$. Since we are
interested in preserving some supersymmetry we will impose the
condition that the action should be invariant under the
transformations (\ref{transf}) for some $\eta$. Varying the above
action, now with $m(x^4)$ a function, gives:
$$
\delta(S) =\int d^5x\,
 m'(x^4) \epsilon_{ij} \epsilon_{ab} {(\tau^3)_c}^b \eta^{\gamma i}
  {(i \Gamma^4 \Gamma^5)_{\gamma}}^{\a} \psi^a_\a \phi^{jc}.
$$
Here $m'(x_4) \equiv dm/dx^4$.
Let us try adding the following term to the Lagrangian:
$$
L_{new}=
{1 \over 4} m'(x^4) \epsilon_{ab} {(\tau^3)_c}^b \epsilon_{ij}
  {(\tau^3)_k}^j  \phi^{ia} \phi^{kc}.
$$
The supersymmetry variation of this term is:
$$
\delta(L_{new}) = \half  m'(x^4) \epsilon_{ab}
{(\tau^3)_c}^b \epsilon_{ij} {(\tau^3)_k}^j  2 \eta^{\a i} \psi^a_\a
\phi^{kc}
$$
We see that this term cancels $\delta(S)$ if
\begin{equation}
{(\tau^3)_j}^i \eta^{\a j} = \eta^{\gamma i}
{(i\Gamma^4 \Gamma^5)_\gamma}^\a   \label{breakhalf}
\end{equation}
This equation breaks half the supersymmetry
and leaves $\SUSY{1}$ in 3+1D.

We thus conclude that a sensible action for a hypermultiplet
with a varying mass is:
\bear
S &=& \int d^5 x
\LeftPar
-{1\over 4} \epsilon_{ij}\epsilon_{ab}
(\partial_{\mu}\phi^{ia} \partial^{\mu}\phi^{jb}
+ m^2 \phi^{ia} \phi^{jb}
- m'(x^4)  {(\tau^3)_c}^b
 {(\tau^3)_k}^j  \phi^{ia} \phi^{kc} ) \nn \\
&&+\half \epsilon_{ab} \psi_\a^a \Gamma^{\mu \a \b}\partial_\u
\psi^b_\b
+ \half i m \epsilon_{ab}{(\tau^3)_c}^b \psi_\a^a
 \Gamma^{5 \a \b}\partial_5
\psi^c_\b
\RightPar.
\label{virkning}
\eear
It preserves the supersymmetry transformations (\ref{transf})
when $\eta$ solves (\ref{breakhalf}).
The equations of motion are:
\bear
\left(\Box + m(x^4)^2\right)\phi^{ia}
- m'(x^4) (\tau^3)^i_j (\tau^3)^a_b \phi^{jb} &=& 0,
\label{dmdx}\\
\Gamma^{\u,{\a\b}}\partial_{\u}\psi_\b^a +
i m(x^4) {(\tau^3)_b}^a \Gamma^{5 \a\b}\psi_\b^b &=& 0.
\nn
\eear

\subsection{Chiral zero modes}
% -----------------------------------------------------------------------
As usual one can reduce the fields along the $x^4$ direction and
find the modes seen from a 3+1D point of view. The $x^4$ direction
is noncompact here. Later we will consider the compactified case.
Let us find the massless modes in 3+1D. Since $\SUSY{1}$
is preserved in 3+1D we know that the fields have to come in
chiral multiplets. The bosons will have a 3+1D massless mode for
every solution of  (setting $y\equiv x^4$):
$$
\left(- {d^2 \over dy^2} + m(y)^2\right) \phi^{ia} 
   -m'(y) {(\tau^3)_b}^a {(\tau^3)_j}^i \phi^{jb} = 0.
$$
The fermions will have zero modes for every solution of:
$$
{(i\Gamma^4 \Gamma^5)_\a}^\b {d \over dy} \psi^a_\b + m
{(\tau^3)_b}^a \psi^b_\a =0.
$$
We see that the bosonic equation is the square of the
fermionic one in a certain sense and that the term
 proportional to $m'(y)$ is essential for this.
The solution to the fermionic equation is:
$$
%\psi(y) ={\mbox{exp}}\left(-\int_0^y m(z)dz (i\Gamma^4 \Gamma^5)\tau^3
%\right)\psi_0
\psi(y) =e^{-i\Gamma^4 \Gamma^5\tau^3 \int_0^y m(y')dy'}
\psi_0.
$$
Here we use matrix notation and suppress indices. Both
matrices $(i\Gamma^4 \Gamma^5)$ and $\tau^3$ have eigenvalues $+1$
and $-1$.
For the solution to be normalizable it is thus necessary
that either:
$$
\int_0^y m(y')dy'  \rightarrow \infty\qquad {\mbox{
for $y \rightarrow \pm \infty$}}.
$$
or
$$
\int_0^y m(y')dy'
\rightarrow - \infty \qquad {\mbox{ for $y \rightarrow \pm
\infty$}}.
$$
In the former case the solution is normalizable if
$\psi_0$ has the same eigenvalue as $(i\Gamma^4 \Gamma^5)$ and
$\tau^3$ and in the latter case the eigenvalues must be opposite.
In both cases we end up with two chiral spinors in 3+1D which are
related by the reality condition (\ref{reality}) leaving one
independent chiral spinor.

The solution to the bosonic equation is:
$$
\phi(y) = e^{-\tau^3_R \tau^3_F\int_0^y m(y')dy'}\phi_0,
$$
where again we suppress indices. The $\tau^3$ matrices are
written with a subindex to distinguish the R-symmetry and
the flavor-symmetry. There is a normalizable solution exactly
in the same two cases of $\int_0^y m(y')dy'$ as above.
In both cases there are two solutions which are related by the
reality condition. So there is one massless complex boson in both
cases. This one pairs up with the chiral fermion to give a
massless $\SUSY{1}$ chiral multiplet as we expect.
(For a similar mechanism, see \cite{KatVaf}.)

The condition on $m(y)$ stated above implies in particular
that $m(y)$ crosses zero at some point. A particular example of
an $m(y)$ that obeys the condition is a function that
goes to $-m_0$ for $y \rightarrow - \infty$, crosses zero
 and goes to $m_0$ for $y \rightarrow \infty$.

\subsection{Flavor current multiplet}\label{flavor}
% -----------------------------------------------------------------------
In subsection (\ref{masshyp})
above, we generated a 4+1D mass by reduction from 5+1D requiring
that the fields have a specific $x^5$ behavior (\ref{reduction}).
If one just compactifies on a circle, the 4+1D theory will have a
tower of Kaluza-Klein states with the lowest one being massless.
The massless mode is the constant mode on the circle. The theory
has a current, $J_\mu$, associated with the $U(1)_F$
 symmetry. We can introduce a background gauge field,
$A_\mu$, that couples to this current.
Creating a Wilson line for the background gauge field, $A_\mu$,
around the circle is equivalent to changing the periodicity
condition of the 5+1D hypermultiplet fields. They will be identified
with themselves up to a $U(1)_F$ rotation. This gives them exactly
the $x^5$ behavior of (\ref{reduction}).  In a circle
compactification with a Wilson line for $A_\mu$ there will still
be a Kaluza-Klein tower of states
in 4+1D but their masses will be shifted
with an amount proportional to the Wilson line.
 The $U(1)_F$ is part of an $SU(2)_F$ symmetry.
The 5+1D hypermultiplet has a current $J_\mu^A$ ($A=1,2,3$
is an index of the $\rep{3}$ of $SU(2)$) associated
with the $SU(2)_F$ flavor symmetry.
This Noether-current is easily found from the action.
By applying supersymmetry transformations to the current one finds
that it is part of the following supermultiplet:
\bear
J_\mu^A &=&
 i{1 \over 4} \epsilon_{ij}\epsilon_{ab}{(\tau^A)_c}^b
   (\phi^{jc}\px{\u}\phi^{ia} -\px{\u}\phi^{jc}\phi^{ia})
 - i \half \epsilon_{ab}{(\tau^A)_c}^b \Gamma_\u^{\a\b}\psi_\a^a\psi_\b^c,
\nn\\
S_\a^{j A} &=&
 i\epsilon_{ba}{(\tau^A)_c}^b \phi^{j a}\psi_\a^c ,
\nn\\
D^{ij A} &=&
  \half i \epsilon_{ba}{(\tau^A)_c}^b \phi^{i c} \phi^{j a}.
\label{JSD}
\eear
Note that $D^{ij A}$ is symmetric in $i$ and $j$.
The SUSY transformations of these operators are:
\bear
\delta J^{\mu A} &=& \epsilon_{ij}\eta^{\a i} {(\Gamma^{\mu \nu})_\a}^\b
                   \partial_\nu S_\b^{jA},
\nn\\
\delta S_\b^{j A} &=& \eta^{j \gamma} \Gamma^\mu_{\b \gamma} J^A_\mu +
                      \epsilon_{ki}\eta^{\gamma k}
                      \Gamma^\mu_{\beta\gamma}\partial_\mu D^{ij A},
\nn\\
\delta D^{ij A} &=& \eta^{\a i}S^{j A}_\a + \eta^{\a j}S^{i A}_\a.
\nn
\eear
In the transformation of $J_\mu^A$ the equation of motion for
$\psi_\a$ was used.

Since a mass in 4+1D comes from the component $A_5$ along the circle,
a mass varying in the $x^4$ direction comes from an $A_5$ which varies
along $x^4$. In other words, there is a nonzero field strength $F_{45}$.
The usual way of coupling $A_\mu$ to a theory is by adding
$$
\int d^6x J_\mu A^\mu
$$
to the action plus a term proportional to $A^2$ in order to
preserve gauge invariance. In the action (\ref{virkning})
the terms proportional to $m$ and $m^2$ come from this
coupling. What about the extra term needed for supersymmetry?
We see that it is proportional to $D^{12 (A=)3}$. Since $m'(x^4)$
is $F_{45}$ we see that the extra term is just proportional to
$$
\int d^6x F_{45}D^{12 (A=)3}.
$$
We will apply these observations to more general systems in the
next section. The important point is that the deformation of
the Lagrangian can be expressed in terms of the current $J_\mu$
and its superpartner $D$ without referring to the specific
fields of the theory.

\section{Construction from 6D}
% =======================================================================
We wish to analyze the situation starting from a general 5+1D theory.
We start with a 5+1D theory with $\SUSY{(1,0)}$ supersymmetry
and a global $U(1)$ symmetry and we
compactify it on $\MT{2}$.
We wish to put a background gauge field $A_\u$ that
is associated to the $U(1)$ symmetry along $\MT{2}$ such that
the first Chern class will be $c_1 = n$.
The question is how do we do it while preserving half the supersymmetry.

\subsection{The current multiplet}
% -----------------------------------------------------------------------
The 5+1D theory has a current $J_\u$ associated with the $U(1)$
symmetry. The current is a member of an $\SUSY{(1,0)}$ multiplet
which also contains a fermionic partner $S^i_\a$ and a bosonic
``D-term'' partner $D^{ij}$ as we saw in subsection (\ref{flavor})
for the free hypermultiplet.
Here, $i,j=1,2$ are $SU(2)_R$ symmetry indices
and $D^{ij}$ is symmetric.
They satisfy:
\bear
\delta J^\mu &=& \epsilon_{ij}\eta^{\a i} {(\Gamma^{\mu \nu})_\a}^\b
                   \partial_\nu S_\b^{j}
\nn\\
\delta S_\b^{j} &=& \eta^{j \gamma} \Gamma^\mu_{\b \gamma} J_\mu +
                      \epsilon_{ki}\eta^{\gamma k} \Gamma^\mu_{\beta
                      \gamma}\partial_\mu(D^{ij})
\label{susyvariation}\\
\delta D^{ij} &=& \eta^{\a i}S^{j}_\a + \eta^{\a j}S^{i}_\a
\nn
\eear
We claim that compactifying on $\MT{2}$ and adding:
\be\label{ajsix}
S_1 = - \int (A_4 J_4 + A_5 J_5 +i F_{45} D^{12} + \cdots)
\ee
to the action gives a supersymmetric theory with $\SUSY{1}$
in the uncompactified 3+1D. The $i$ in the second term is
necessary to make the
action real, since $D^{12}$ is imaginary.
The $(\cdots)$ represent $O(A_\u^2)$ terms that are dictated by
$U(1)$ gauge invariance. For example,
if under a local $U(1)$ transformation
$$
\delta J_\u = \px{\u}\epsilon \Theta,
$$
we have to add $\half A_\u A^\u \Theta$ to the Lagrangian.

In order to see that $\SUSY{1}$ is unbroken we calculate the
supersymmetry variation of $S_1$ using (\ref{susyvariation}).
\bear
\delta S_1 &=&
\int A_\mu(\epsilon_{ij} \eta^{\a i} {(\Gamma^{\mu\nu})_\a}^\b
 \partial_\nu S^j_\b )
  + i F_{45} ( \eta^{\a 1}S^{2}_\a + \eta^{\a 2}S^{1}_\a ) \nn \\
&=& \int (F_{45}{(\Gamma^{45})_\a}^\b \eta^{\a 2} +iF_{45}
    \eta^{\b 2})S^1_\b +
      (- F_{45}{(\Gamma^{45})_\a}^\b \eta^{\a 1} +iF_{45}
    \eta^{\b 1})S^2_\b
\nn
\eear
which is equal to zero if
$$
 {(\Gamma^{45})_\a}^\b \eta^{\a 1} = i \eta^{\b 1},\qquad
 {(\Gamma^{45})_\a}^\b \eta^{\a 2} = -i \eta^{\b 2}.
$$
These two equations are complex conjugate of each other.
We see that we are left with $\SUSY{1}$ in 3+1D.

\subsection{Example -- a free hypermultipet}
% -----------------------------------------------------------------------
After compactification on a $\MT{2}$ to 3+1D we would like to know
the masses of the fields. There will be a Kaluze-Klein tower of
fields. In the low energy limit we are, of course, only interested
in the massless fields.
    Let us go back to the free hypermultiplet and calculate
the Kaluza-Klein masses. We need only do it for the fermions
because of $\SUSY{1}$. The Dirac equation for the fermions reads
$$ \Gamma^\mu \nabla_\mu \psi = 0 $$ where $\nabla_\mu =
\partial_\mu + i A_\mu$ is the covariant derivative with respect
to the $U(1)$ symmetry. In our case the only nonzero components of
$A_\mu$ are $A_4$ and $A_5$. In reducing to 3+1D $\psi$ can be
written as $$ \psi = \psi_L \phi_L + \psi_R \phi_R $$ where
$\psi_L,\psi_R$ are left- and righthanded spinors in 3+1D and
$\phi_L,\phi_R$ are left- and righthanded spinors on $\MT{2}$.
Plugging into the Dirac equation we get the following formula for
the mass $m$ in 3+1D. \bear ( \nabla_4 \Gamma_4 + \nabla_5
\Gamma_5) \phi_R &=& m \phi_L \nn \\ ( \nabla_4 \Gamma_4 +
\nabla_5 \Gamma_5) \phi_L &=& - m^* \phi_R \label{vh} \eear The
mass $m$ is a complex number. The physical mass is the absolute
value of $m$. The phase can be transformed away by redefining
$\phi_L$, say. The phase would then show up in the couplings. In
the free theory there is no meaning to them. We will just rotate
the phase away for now and let $m$ be real. We see that for $m
\neq 0$, $\phi_L$ and $\phi_R$ come in pairs. This implies that in
3+1D $\psi_L$ and $\psi_R$ come in pairs of the same mass. This is
as it should be, since a chiral spinor that is charged
under a $U(1)$ symmetry cannot be
massive. Both a lefthanded and a righthanded spinor are needed for
a mass term.
    However for $m=0$ there is no relation between a
lefthanded solution and a righthanded one. For each solution of
$$
( \nabla_4 \Gamma_4 + \nabla_5 \Gamma_5) \phi_R =0
$$
there is a massless righthanded fermion in 3+1D and
for each solution of
$$
( \nabla_4 \Gamma_4 + \nabla_5 \Gamma_5) \phi_L =0
$$
there is a massless lefthanded fermion in 3+1D.

Eq. (\ref{vh}) implies
second order differential equations for $\phi_L$ and $\phi_R$:
\bear
(\nabla_4^2 + \nabla_5^2 - F_{45}) \phi_L &=& -m^2 \phi_L \nn \\
(\nabla_4^2 + \nabla_5^2 + F_{45}) \phi_R &=& -m^2 \phi_R \label{ligninger}
\eear
These equations are the same as the ones determining the
boson masses. It had to be so due to the supersymmetry.
In these equations $A_\mu$ is a connection in a $U(1)$-bundle over
$\MT{2}$ and $\phi_{L,R}$ are sections of this circle-bundle.
The setup here is the same as a charged particle on a torus moving
in a background magnetic field (Landau levels).
For a general $A_\mu$ the eigenvalues $m$ are not known, to our knowledge.

We can say more about the case of $m=0$. Here we find the
zero modes of the Dirac equation in 2 dimensions for respectively lefthanded
and righthanded spinors. The number of those will depend on the
gauge field $A_\mu$ but the difference between the number of lefthanded
and righthanded zero modes is known as the index of the Dirac operator.
It is equal to the first chern class, $c_1$, of the circle-bundle.
$$
c_1 = {1 \over 2\pi} \int_\MT{2} F_{45}
$$
For a generic gauge field there will be
$|c_1|$ solutions of one kind and 0 of the other kind. But for special
gauge fields it could be different. An example of a special case
is the case of $A_\mu =0$. Here $c_1=0$. There is one zero mode of
each chirality, namely the constant function.
    We thus conclude that in the theories under consideration
the hypermultiplets will give rise to
$c_1$ massless chiral multiplets. Even in the special cases mentioned
above this will also be the case, since the couplings generically
will lift the accidental pairs and still leave us with $c_1$ massless
 chiral multiplets.

Now we will consider the special case of constant $F_{45}$, where
the problem has an explicit solution. Let the first chern class be
$c_1=n$. We will take the fields to obey the following boundary
conditions.
\bear
\phi(x_4,x_5+2\pi R_5) &=& \phi(x_4,x_5 ) \nn \\
\phi(x_4 + 2\pi R_4, x_5) &=& e^{-i n{{x_5}\over {R_5}}}\phi(x_4,x_5) \nn
\eear
Here $\phi$ denotes both $\phi_R$ and $\phi_L$. The gauge field can
be gauge transformed to the following form:
\bear
A_4(x_4,x_5) &=& a_4 \nn \\
A_5(x_4,x_5) &=& {{n x_4} \over {2\pi R_4 R_5}} + a_5 \nn
\eear
Here $a_4,a_5$ are constants. On the plane they could be gauged to
zero, but on the torus they are there in general. The eigenvalue
equations (\ref{ligninger}) now read
\bear
\left\lbrack
\left(\partial_4 + i a_4\right)^2 
+\left(\partial_5 +i{nx_4 \over 2\pi R_4 R_5} + ia_5\right)^2
\pm F_{45}\right\rbrack\phi &=& -m^2 \phi,
\nn
\eear
where the $\pm$ refers to $\phi_R$ and $\phi_L$, respectively. The
periodicity conditions above imply that we can write $\phi$ as:
\be
\phi(x_4,x_5) = \sum_{k= - \infty}^{\infty} e^{ik{x_5 \over R_5}}
\phi_k(x_4)    \qquad 0 \le x_4 \le 2\pi R_4,  \label{fourier}
\ee
with the boundary condition:
\be
\phi_k(2\pi R_4) = \phi_{k+n}(0). \label{boundcond}
\ee
The equation for $\phi_k$ becomes
\bear
\left\lbrack \left(\partial_4 + i a_4\right)^2 + \left(i
{k \over R_5} +i{nx_4 \over 2\pi R_4 R_5} + ia_5\right)^2 \pm
F_{45}\right\rbrack\phi_k(x_4) &=& -m^2 \phi_k(x_4),
\qquad 0 \le x_4 \le 2\pi R_4
\nn\\ &&
\label{lignmode}
\eear
Using the boundary condition
(\ref{boundcond}) we can define $n$ functions, $f_k,
k=0,1,...,n-1$ on the real line:
$$
f_k(x_4) = \phi_{k+ln}(x_4 -  2\pi R_4 l)
\qquad {\mbox{for $2\pi R_4 l\le x_4 \le 2\pi R_4 (l+1)$}}.
$$
It follows from (\ref{lignmode}) that $f_k$ obeys
\be
\left\lbrack
\left(\partial_4 + i a_4\right)^2 
+\left(i {k \over R_5} +i{nx_4 \over 2\pi R_4
R_5} + ia_5\right)^2 \pm F_{45}\right\rbrack f_k(x_4) = -m^2 f_k(x_4),
\qquad -\infty < x_4 < \infty. \label{lignf}
\ee
Here $k=0,1,...,n-1$ and
$\pm$ still refers to the two chiralities. We are only interested
in normalizable solutions. The norm square of a field $\phi$ in
(\ref{fourier}) is equal to the sum of the norm squares of the $n$
functions on the real line, $f_k$. This means that the eigenvalues
and eigenfunctions are exactly the normalizable solutions to
(\ref{lignf}).

To solve (\ref{lignf}) we first redefine $f_k$ by a phase to set
$a_4$ to zero. This can now be done since $x_4$ runs over the real
line. The equation becomes the eigenvalue problem for a one
dimensional harmonic oscillator. The eigenvalues are:
$$
m_j^2 = (j+\half \mp \half) {n \over \pi R_4 R_5} \qquad j=0,1,2,\dots
$$
for each $k=0,1,...,n-1$. We see that there is a n-fold degeneracy
of all masses. There are $n$ massless modes of one chirality and
zero of the other. For the massive levels there is an equal number
of solutions of each chirality. These features were general as
discussed above and it is nice to see how it works in the special
case of constant $F_{45}$.

We thus conclude that the free hypermultiplet compactified in this
way produces
 $n$ chiral multiplets with zero mass as well as a tower
of nonchiral (double) multiplets $\Phi_j^{k,\pm}$ ($j=1,\dots$ and
$k=0,1,\dots,n-1$) with masses,
$$
m_j^2 = {{j n}\over {\pi R_4 R_5}}.
$$
The superpotential therefore contains a term,
$$
\sum_{k=1}^n\sum_{j=1}^\infty
\left({{j n}\over {\pi R_4 R_5}}\right)^{1/2}
\Phi_j^{k,+}\Phi_j^{k,-}.
$$

\subsection{$\sigma$-models}
% -----------------------------------------------------------------------
\def\bpartial{{\overline{\partial}}}
\def\bphi{{\overline{\phi}}}
\def\bj{{\bar{j}}}
\def\bk{{\bar{k}}}
\def\bl{{\bar{l}}}
\def\bz{{\bar{z}}}
\def\bD{{\overline{D}}}
\def\wOmega{{\widetilde{\Omega}}}

The previous example can be generalized to $q$ hypermultiplets
describing a low-energy $\sigma$-model with a hyper-K\"ahler
target space, ${\cal M}$, of dimension $4q$.
Let us also assume that ${\cal M}$ has a $U(1)$ isometry that
is related to a hyper-K\"ahler moment map.
Recall that a hyper-K\"ahler manifold has a $\CP{1}$-family
of complex structures and each complex structure has
its own K\"ahler class.
The collection of K\"ahler 2-forms can be written as:
$$
\omega = \sum_{a=1}^3 c_a\omega_a,\qquad \sum c_a^2 = 1.
$$
Here, the $\omega_a$'s are (real) 2-forms and the $c_a$'s are real 
coefficients. They satisfy,
$$
g_{IK}\omega_a^{IJ}\omega_b^{KL}+ g_{IK}\omega_b^{IJ}\omega_a^{KL}
 = 2\delta_{ab} g^{JL},
$$
where $g_{IJ}$ is the metric ($I,J,K=1\dots 4q$).
A hyper-K\"ahler moment map is a $\CP{1}$-family of functions
on ${\cal M}$:
$$
\mu = \sum_{a=1}^3 c_a\mu_a.
$$
They satisfy,
$$
\omega_a^{IK}\px{K}\mu_b +\omega_b^{IJ}\px{J}\mu_a
= 2\delta_{ab}\xi^I,
$$
where $\xi^I$ is the Killing vector for the $U(1)$ isometry.

Now, let us consider a 5+1D $\sigma$-model with target space
${\cal M}$ (the hypermultiplet moduli space).
(See \cite{BagWit} and \cite{HKLR}.)
The $U(1)$ current is given by:
$$
J_\mu = \xi_I\px{\mu}\phi^I.
$$
The role of the triplet of operators $D^{ij}$ from (\ref{susyvariation})
is played by the triplet of moment maps $\mu_a$ ($a=1\dots 3$).
When we compactify on $\MT{2}$, (\ref{ajsix}) becomes:
\be\label{ajmu}
S_1 = - \int (A_4 J_4 + A_5 J_5 +i F_{45} \mu_1 + \cdots)
\ee
Let us discuss the low-energy description of this model.
We wish to find the dimension of the moduli space of solutions
to the scalar equations of motion.
The kinetic part of the $\sigma$-model:
$$
\int g_{i\bj}(\phi,\bphi)\bpartial\phi^i\partial\bphi^\bj
+\int g_{i\bj}(\phi,\bphi)\partial\phi^i\bpartial\bphi^\bj,
$$
leads to the following equations of motion:
\bear
0 &=&
-\partial(g_{i\bj}\bpartial\phi^i)
-\bpartial(g_{i\bj}\partial\phi^i)
+\px{\bj}g_{i\bk}\partial\bphi^\bk\bpartial\phi^i
+\px{\bj}g_{i\bk}\bpartial\bphi^\bk\partial\phi^i.\nn
\eear
We use the K\"ahler condition:
$$
\px{\bj}g_{i\bk}=\px{\bk}g_{i\bj} = g_{i\bl}\Gamma^\bl_{\bj\bk}
$$
and obtain:
%\bear
%0 &=&
%-\partial(g_{i\bj}\bpartial\phi^i)
%-\bpartial(g_{i\bj}\partial\phi^i)
%+\px{\bj}g_{i\bk}\partial\bphi^\bk\bpartial\phi^i
%+\px{\bj}g_{i\bk}\bpartial\bphi^\bk\partial\phi^i
%\nn\\ &=&
%-2g_{i\bj}\partial\bpartial\phi^i
%-\px{k}g_{i\bj}\partial\phi^k\bpartial\phi^i
%-\px{\bk}g_{i\bj}\partial\bphi^\bk\bpartial\phi^i
%\nn\\ &&
%-\px{k}g_{i\bj}\bpartial\phi^k\partial\phi^i
%-\px{\bk}g_{i\bj}\bpartial\bphi^\bk\partial\phi^i
%\nn\\ &&
%+\px{\bj}g_{i\bk}\partial\bphi^\bk\bpartial\phi^i
%+\px{\bj}g_{i\bk}\bpartial\bphi^\bk\partial\phi^i
%\nn\\ 
%&=&
%-2g_{i\bj}\partial\bpartial\phi^i
%-\px{k}g_{i\bj}\bpartial\phi^k\partial\phi^i
%-\px{k}g_{i\bj}\partial\phi^k\bpartial\phi^i
%\nn\\
%&=&
%-2g_{i\bj}\partial\bpartial\phi^i
%-g_{l\bj}\Gamma^l_{ik}\bpartial\phi^k\partial\phi^i
%-g_{l\bj}\Gamma^l_{ik}\partial\phi^k\bpartial\phi^i
%\nn\\
%&=&
%-2g_{i\bj}\partial\bpartial\phi^i
%-2g_{i\bj}\Gamma^i_{kl}\partial\phi^k\bpartial\phi^l
%\eear
$$
(D\bD\phi)^i = 0,\qquad \bD D\bphi^\bj = 0.
$$
Here $D$ is the covariant derivative:
$$
(\bD\phi)^i = \bpartial\phi^i,\qquad
(D\bD\phi)^i = \partial\bpartial\phi^i 
     +\Gamma^i_{jk}\partial\phi^j\bpartial\phi^k.
$$
This implies:
$$
\bpartial\phi^i = 0,\qquad \partial\bphi^\bj = 0.
$$
The zero modes are thus holomorphic curves from $\MT{2}$ into
the target space, as is well known.
To incorporate the gauge field $A_\u$ we replace $\partial$
and $\bpartial$ with the $U(1)$-covariant derivative:
$$
(\bD\phi)^j = \bpartial_\bz\phi^j - i A_\bz\xi^j.
$$
Now let us fix the complex structure that corresponds to
$\omega_1$ (out of the 3 $\omega_a$'s).
We can then express the Killing vector, $\xi^j$, in terms of
$\mu_1$ as:
\be\label{ximu}
\xi^j = g^{j\bk}\px{\bk}\mu_1.
\ee
The zero modes corresponding to  (\ref{ajmu}) are easily seen to
satisfy:
\be\label{eqmo}
0 = \bpartial_\bz\phi^j - i A_\bz\xi^j.
\ee
How many zero modes do we get?
Let us assume that $\phi^j$ is a solution and study the linearized
equation:
$$
0 = \bpartial_\bz\delta\phi^j - i A_\bz\px{k}\xi^j \delta\phi^k
  -i A_\bz\px{\bk}\xi^j\delta\bphi^\bk.
$$
Using (\ref{ximu}) we see that:
$$
\px{\bk}\xi_\bl = \px{\bl}\xi_\bk,
$$
but since $\xi$ is assumed to be a Killing vector it must satisfy:
$$
\px{\bk}\xi_\bl +\px{\bl}\xi_\bk = 2\Gamma^\bj_{\bk\bl}\xi_\bj
$$
so
$$
\px{\bk}\xi_\bl = \Gamma^\bj_{\bk\bl}\xi_\bj
$$
Also,
$$
\px{\bk}g^{j\bl} = -g^{j\bar{n}}\px{\bk}g_{m\bar{n}}g^{m\bl}
  = -g^{j\bar{n}}\Gamma^{\bl}_{\bk\bar{n}}
$$
Therefore,
$$
\px{\bk}\xi^j = 0.
$$
The linearized equations of motion are therefore:
$$
0 = \bpartial_\bz\delta\phi^j - i A_\bz\px{k}\xi^j \delta\phi^k.
$$
To solve this we write the $2q\times 2q$ matrix with elements
$A_\bz\px{k}\xi^j$ as:
$$
-i A_\bz\px{k}\xi^j = (\Omega^{-1})_k^l\bpartial_\bz\Omega^j_l,
$$
where $\Omega(z,\bz)\in GL(2q,\BC)$.
We find that:
$$
\bpartial_\bz(\Omega^j_k \delta\phi^k) = 0.
$$
Thus $\Omega\delta\phi$ is a holomorphic section of a vector-bundle.
Moreover, from the Killing vector equation:
$$
\px{\bk}\xi_l +\px{l}\xi_\bk = 
 2\Gamma^\bj_{\bk l}\xi_\bj+2\Gamma^j_{\bk l}\xi_j = 0.
$$
We therefore find:
$$
\px{l}\xi^i = \px{l}g^{i\bk}\xi_\bk +g^{i\bk}\px{l}\xi_\bk
 = -\Gamma^i_{lk}\xi^k - g^{i\bk}\px{\bk}\xi_l
$$
Using (\ref{eqmo}) we can write:
$$
(\Omega^{-1})_k^l\bpartial_\bz\Omega^j_l =
-\Gamma^j_{kl}\bpartial_\bz\phi^l 
  +i A_\bz g^{j\bl}\px{\bl}\xi_k
$$
Now $\delta\phi^j$ is a section of the pullback $\phi^* T{\cal M}$
of the tangent-bundle $T{\cal M}$ of ${\cal M}$ under the map
$\phi:\MT{2}\mapsto {\cal M}$.
This vector-bundle has the connection
$\Gamma^j_{kl}\bpartial_\bz\phi^l$.
Thus,
the vector-bundle $V$, of which $\Omega\delta\phi$ is a holomorphic
section can be described as follows.
Find $\wOmega\in GL(2q,\BC)$ such that:
$$
(\wOmega^{-1})_k^l\bpartial_\bz\wOmega^j_l =
 i A_\bz g^{j\bl}\px{\bl}\xi_k 
 = i A_\bz g^{j\bl}\px{\bl}\px{k}\mu_1.
$$
Then, $\wOmega$ is a section of a principal bundle
with the same structure group as $V$.
This means the following:
Let $\MT{2}$ be described by $z$, as we did,
with
$$
z\sim z+1,\qquad z\sim z+\tau.
$$
If $s$ is a section of $V$ then the boundary conditions on $s$ are
that $\wOmega(z,\bz)^{-1} s$ should be continuous.

The eigenvalues of the $GL(2q,\BC)$ 
matrix with elements $g^{j\bl}\px{\bl}\xi_k$ pulled back 
to $\MT{2}$ are constants, and therefore also integers.
The fact that the invariant polynomial
$P(\lambda)\equiv\det (g^{j\bl}\px{\bl}\xi_k-\lambda \delta^j_k)$
is constant follows from $\partial_{\bk}\xi^l = 0$. It implies
that $\partial_{\bk} P(\lambda) = 0$. Thus $P(\lambda)$ is a holomorphic
function. If ${\cal M}$ were compact this is enough.
Even if it is not compact,
it still follows that the pullback of $P(\lambda)$
to $\MT{2}$ is holomorphic and therefore constant.
Thus, the vector-bundle $V$ splits into a product:
$\bigotimes_{i=1}^{2q} {\cal O}(n\lambda_i)$ where $\lambda_i$ are
the eigenvalues of $P(\lambda)$. They must therefore be integers.

% - - - - - - - - - - - - - - - - - - - - - - - - - - - - - - - - - - - -
%%% ------------------------- CUT HERE ----------------------------------

\subsection{Coupling to a vector multiplet}
% -----------------------------------------------------------------------
Now let us start with a 5+1D hypermultiplet in the representation
$\rep{N}$ ($\rep{\bar{N}}$)
of $SU(N)$ and couple it to a 5+1D $SU(N)$ vector-multiplet.
Although this  is a nonrenormalizable interaction, we can think
of it as the low-energy description of a sector of one of
the little-string theories of \cite{IntNEW}.
The 5+1D coupling of the vector-multiplet to the hypermultiplet
preserves $SU(2)_R\times U(1)_F$. Out of the two chiral fermions
$\psi^a_\a$ ($a=1,2$) one transforms in the $\rep{N}$ of $SU(N)$
and the other transforms in the $\rep{\bar{N}}$ of $SU(N)$.

Let us classically reduce, as before, on $\MT{2}$ with a global
$U(1)$ background field with first Chern class $c_1 = n$.
The hypermultiplet gives rise to $n$ chiral multiplets
$\Phi_0^{(k),+}$ ($k=1\dots n$) in the $\rep{N}$ of $SU(N)$ as
well as a tower of massive multiplets
$\Phi_j^{(k),\pm}$ ($j=1\dots$) where $\Phi_j^{(k),+}$ is in the
$\rep{N}$ of $SU(N)$ and $\Phi_j^{(k),-}$ is in the $\rep{\bar{N}}$.
Their masses are given by the superpotential,
$$
\sum_{k=1}^n\sum_{j=1}^\infty
  \left({{j n}\over {\pi R_4 R_5}}\right)^{1/2}
  \Phi_j^{(k),+}\Phi_j^{(k),-}.
$$
The 5+1D vector-multiplet gives rise to an $\SUSY{1}$ vector-multiplet
in 3+1D and a chiral multiplet $\Phi_{ad}$
in the adjoint representation of $SU(N)$.
There is also a Yukawa coupling of the fields
$\Phi_{ad}$, $\Phi_{j+1}^{(k),-}$ and $\Phi_j^{(k),+}$.
%%% The Yukawa coupling in the superpotential is given by:
%%% $$
%%% \sum_{k=1}^n\sum_{j=0}^\infty
%%%   (j+1)^{1/2}(\cdots)
%%%   \Phi_j^{(k),+}\Phi_{ad}\Phi_{j+1}^{(k),-}.
%%% $$

% - - - - - - - - - - - - - - - - - - - - - - - - - - - - - - - - - - - -
%%% ------------------------- CUT HERE ----------------------------------

%\vskip 0.5cm
%{\bf this is file ``lowe.tex''}
%\vskip 0.5cm

\section{Compactifying the BI theory}
% =======================================================================
We will now construct a specific example that produces
chiral matter in 3+1D by compactifying the
Blum-Intriligator (BI) theories \cite{BluInt}.

\subsection{Preliminaries}
% -----------------------------------------------------------------------
Compactifying the BI theory of $N$ M5-branes at an $A_{k-1}$
singularity on $\MS{1}$ of radius $R$ one obtains a low-energy
description given by a gauge theory with gauge group
$$
SU(N)_1\times SU(N)_2\times \cdots\times SU(N)_k.
$$
The sub-indices are added for purposes of identification. There are also
hypermultiplets in the $(\overline{N}_i,N_{i+1})$ representation
(with $k+1\equiv 1$). On top of that there are $(k-1)$ more $U(1)$
vector multiplets. The scalar components set the coupling
constants of the $k$ $SU(N)$ gauge groups. These coupling
constants, $g_i$, ($i=1\dots k$) satisfy $$ \sum_{i=1}^k {1\over
{g_i^2}} = {1\over R}. $$ If we compactify on another $\MS{1}$ of
radius $R'$ we obtain a 3+1D gauge theory at low-energies. The
$(k-1)$ $U(1)$ vector multiplets that set the gauge couplings
decouple and the gauge couplings become background parameters.
The interacting gauge theory has a gauge group $SU(N)^k$
and $(\overline{N}_i,N_{i+1})$ hypermultiplets. The coupling constants
and $\theta$-angles are set by $(k-1)$ background parameters
(originating from the original $(k-1)$ $U(1)$ vector multiplets)
and subject to the condition that
$$
\sum_{i=1}^k \tau_i = i
{{R'}\over {R}},\qquad \tau_i \equiv {{\theta_i}\over {2\pi}} +
{{8\pi i}\over {g_i^2}}
$$

\subsection{Adding the background $U(1)$ field}\label{addb}
% -----------------------------------------------------------------------
Now we take a specific 5+1D theory -- the BI theory.
Also, let the complex structure $\tau$ of $\MT{2}$ become very large.
We can take $\MT{2}=\MS{1}\times\MS{1}$ with one $\MS{1}$ of radius
$R_4$ and the other with radius $R_5\ll R_4$.
We can first reduce the theory along $R_5$. The holonomy
$W(x_4) = \int_0^{2\pi R_5} A_5 (x_4, x_5) dx_5$ varies from
$0$ to $2\pi n$ as $x_4$ varies from $0$ to $2\pi R_4$.

For a fixed $x_4$, the reduction of the BI theory along $\MS{1}$ with
Wilson line $W(x_4)$ was studied in \cite{CGK,CGKM}.
For small $W(x_4)$ and at low energies $0\le E \ll R_5^{-1}$
the theory is described by an effective 4+1D Lagrangian which is
the quiver theory of \cite{DouMoo} of $N$ D4-branes at an $A_{k-1}$
singularity but such that the hypermultiplets have a mass
$m = W(x_4) R_4^{-1}$.
For generic $x_4$ the mass is of the order of $R_4^{-1}$.
There are $n$ values of $x_4$ for which $W(x_4)$ is a multiple of $2\pi$
and in the vicinity of those points the mass $m$ varies from a small
negative to a small positive value.
According to the discussion in subsection~(\ref{varmass}),
the 3+1D low-energy description contains a chiral multiplet for
every time the mass crosses zero.
Note that the term $F_{56} D^{12}$ in (\ref{ajsix}) becomes the term
proportional to $dm/dx^4$ in (\ref{dmdx}).
In subsection~(\ref{varmass}) the $4^{th}$ direction (counting from
$0\dots 4$) was infinite and
there was a continuum of massive modes with arbitrarily low mass.
In our case the $4^{th}$ direction is compact and therefore we expect
a discrete spectrum with the first level of order $R_4^{-1}$.
The chiral mode is likely to remain massless because of arguments
similar to those of \cite{WitINDEX}.

The low-energy description in 3+1D will therefore contain $n$
chiral multiplets for each hypermultiplet of the quiver theory.
We obtain an $SU(N)^k$ vector
multiplets of $\SUSY{2}$ supersymmetry together with $n$ copies
of chiral multiplets (of $\SUSY{1}$ supersymmetry) in the 
$(\overline{N}_i,N_{i+1})$ representations, for each $i=1\dots k$.
The $\SUSY{2}$ vector multiplets should be decomposed into
$\SUSY{1}$ vector multiplets and chiral multiplets in the adjoint
representation of the fields.

Let us now discuss the issue of whether the adjoint multiplets
have a superpotential or not.
On the face of it, the adjoint mutliplets can receive a mass term.
In the limit that we have been using, $R_4\gg R_5$, the mass term,
if it exists, might be of the order of $R_5^{-1}$.
However, the 6D origin of the expectation value of the
chiral multiplets is the expectation values for the $k(N-1)$
tensor multiplets of the 6D theory.
Specifically, let $\Phi$ be the scalar  of one of those tensor
multiplets and let $B_{45}$ be the component
of the anti-self-dual tensor field
corresponding to it.
We can set $\phi = 4\pi^2 (\Phi + i B_{45}) R_4 R_5$.
In the limit that $\Phi R_4 R_5$ is large,
we can trust the 6D low-energy description
of the Coulomb branch of the BI theory and dimensionally reduce
the 6D low-energy effective action to 4D on $\MT{2}$ with twists.
Because of the periodicity $\phi\sim \phi +2\pi i$,
a superpotential for $\phi$ has to have the form
$\sum a_n e^{-n\phi}$. We recognize this as the contribution
of instantons made from strings of the 6D BI-theory wrapped on $\MT{2}$.
To determine whether such instantons contribute to the superpotential
we have to count the zero modes of the fermions in the low-energy
effective action that describes the world-sheet of the string.
%%% These fermions are charged under the $U(1)$ global symmetry and
%%% each fermion will therefore have $c_1=n$ zero modes on the $\MT{2}$.
%%% The fermions that are left-moving in 3+1D are charged under the
%%% $U(1)$ while the 3+1D right-movers are 
%%% However, these fermions come in multiplets of the $SU(2)_R$ symmetry.
%%% Since the $U(1)_$ bundle over the $\MT{2}$ doesn't break
%%% the $SU(2)_R$ symmetry we get an even number of zero modes
The world-sheet theory that lives on the string of the BI theory
can be deduced by dimensionally reducing
the theory that lives on the M2-brane
and an $A_{k-1}$ singularity on a segment between two M5-branes,
setting the boundary conditions appropriately.
It seems that the 1+1D effective theory always has a supermultiplet
of $\SUSY{(2,2)}$ supersymmetry which comprises of 4 scalars
(describing transverse motion of the string inside the 5+1D
space) and fermions that are uncharged under $U(1)$.
Because they are uncharged, and because it is only the interaction
with this global $U(1)$ that breaks the supersymmetry into $\SUSY{1}$
in 3+1D, the instanton will have twice as many fermionic zero modes
than required for a superpotential. It will therefore not
contribute to a superpotential.

\section{Discussion}
% =======================================================================
We argued that chiral gauge theories can be realized as a low-energy
limit of certain compactifications of 6D conformal field theories.
There are several issues that we have not addressed in this paper.
In section~(\ref{addb}) we argued that the particular compactification
of the BI theory that we studied gives an $SU(N)^k$ gauge vector
multiplets of $\SUSY{2}$ supersymmetry together with $n$ copies
of chiral multiplets (of $\SUSY{1}$ supersymmetry) in the 
$(\overline{N}_i,N_{i+1})$ representations, for each $i=1\dots k$.
Some questions for further study would be:
\begin{itemize}
\item
Do the adjoint chiral multiplets get a mass term?
\item
Can we realize the compactifications in an M-theory setting?
That is, can we find a supergravity solution with M5-brane
whose low-energy is described by the compactifications we considered?
\item
In that case,
are these models dual to other chiral gauge field constructions
similar to those in
 \cite{LPTi,LPTii,HanZaf} or chiral F-theory compactifications
\cite{Curio} (and see also \cite{Ovrut} and refs. therein)?
Are they dual to the new compactifications discovered
in \cite{DRS}?\footnote{We are grateful to S. Sethi for discussions
on this point.}

\end{itemize}

%%% ------------------------- CUT HERE ---------------------------------%

%=======================================================================%
% Acknowledgments
%=======================================================================%
%\bigbreak\bigskip\bigskip
%\centerline{\bf Acknowledgments}\nobreak
\section*{Acknowledgments}
We  have benefited from various discussions with
Sav Sethi and Ashvin Vishwanath on issues related to this paper.
We are also grateful to Ken Intriligator for discussions
and explanations about the BI thoeries.
The research of C.S.C was supported by National Science Foundation
graduate fellowship.
The research of O.J.G was supported by National Science Foundation grant
No. PHY98-02484.
The research of M.K. was supported by National Science Foundation grant
No. PHY94-07194.

%%% ------------------------- CUT HERE ---------------------------------%

%%%%%%%%%%%%%%%%%%%%%%%%%%%%%%%%%%%%%%%%%%%%%%%%%%%%%%%%%%%%%%%%%%%%
%  B I B L I O G R A P H Y                                         %
%%%%%%%%%%%%%%%%%%%%%%%%%%%%%%%%%%%%%%%%%%%%%%%%%%%%%%%%%%%%%%%%%%%%
%%% \documentstyle[12pt]{article}
\def\np#1#2#3{{\it Nucl.\ Phys.} {\bf B#1} (#2) #3}
\def\pl#1#2#3{{\it Phys.\ Lett.} {\bf B#1} (#2) #3}
\def\physrev#1#2#3{{\it Phys.\ Rev.\ Lett.} {\bf #1} (#2) #3}
\def\prd#1#2#3{{\it Phys.\ Rev.} {\bf D#1} (#2) #3}
\def\ap#1#2#3{{\it Ann.\ Phys.} {\bf #1} (#2) #3}
\def\ppt#1#2#3{{\it Phys.\ Rep.} {\bf #1} (#2) #3}
\def\rmp#1#2#3{{\it Rev.\ Mod.\ Phys.} {\bf #1} (#2) #3}
\def\cmp#1#2#3{{\it Comm.\ Math.\ Phys.} {\bf #1} (#2) #3}
\def\mpla#1#2#3{{\it Mod.\ Phys.\ Lett.} {\bf #1} (#2) #3}
\def\jhep#1#2#3{{\it JHEP.} {\bf #1} (#2) #3}
\def\atmp#1#2#3{{\it Adv.\ Theor.\ Math.\ Phys.} {\bf #1} (#2) #3}
\def\jgp#1#2#3{{\it J.\ Geom.\ Phys.} {\bf #1} (#2) #3}
\def\hepth#1{{\it hep-th/{#1}}}

%%% \begin{document}

%%% \end{document}
\end{document}